\documentclass[aps,pre,twocolumn,superscriptaddress,noshowpacs,longbibliography]{revtex4-2}
\usepackage{amsmath}
\usepackage{amssymb}
\usepackage{tikz}
\usepackage{graphicx}
\usepackage{dcolumn}
\usepackage{bm}
\usepackage{epsfig}
\usepackage{psfrag}
\usepackage{latexsym}
\def\II{\hbox{$1\hskip -1.2pt\vrule depth 0pt height 1.6ex width 0.7pt\vrule depth 0pt height 0.3pt width 0.12em$}}
\newcommand{\reffig}[1]{\mbox{Fig.~\ref{#1}}}
\newcommand{\refeq}[1]{\mbox{Eq.~(\ref{#1})}}
\newcommand{\refsec}[1]{\mbox{Sec.~\ref{#1}}}

\newcommand{\be}{\begin{equation}}
\newcommand{\ee}{\end{equation}}
\newcommand{\bal}{\begin{align}}
\newcommand{\eal}{\end{align}}
\newcommand{\ba}{\begin{eqnarray}}
\newcommand{\ea}{\end{eqnarray}}
\newcommand{\T}{${\mathcal T}\,$}

\usepackage{dsfont}
\usepackage{color}
\usepackage{float}
\usepackage{ulem}
\usepackage{soul}

\usepackage{xcolor}
\usepackage{url}
\begin{document}
\title{Experimental study of the distributions of off-diagonal scattering-matrix elements of quantum graphs with symplectic symmetry}
\author{Jiongning Che}\email{chejn20@gmail.com}
\affiliation{Lanzhou Center for Theoretical Physics, Lanzhou University, Lanzhou, Gansu, China}
\affiliation{Yangtze Delta Region Institute, University of Electronic Science and Technology of China, Huzhou, China}
\author{Nils Gluth}\email{nils.gluth@stud.uni-due.de}
\affiliation{Fakult\"at f\"ur Physik, Universit\"at Duisburg-Essen, Duisburg, Germany}
\author{Simon K\"ohnes}\email{simon.koehnes@uni-due.de}
\affiliation{Fakult\"at f\"ur Physik, Universit\"at Duisburg-Essen, Duisburg, Germany}
\author{Thomas Guhr}\thanks{Corresponding author: thomas.guhr@uni-due.de}
\affiliation{Fakult\"at f\"ur Physik, Universit\"at Duisburg-Essen, Duisburg, Germany}
\author{Barbara Dietz}\thanks{Corresponding author: bdietzp@gmail.com}
\affiliation{Lanzhou Center for Theoretical Physics, Lanzhou University, Lanzhou, Gansu, China}
\affiliation{Center for Theoretical Physics of Complex Systems, Institute for Basic Science, and Basic Science Program, Korea University of Science and Technology (UST), Daejeon, Korea}

\begin{abstract}
	We report on experimental studies of the distribution of the off-diagonal elements of the scattering ($S$) matrix of open microwave networks with symplectic symmetry and a chaotic wave dynamics. These consist of two geometrically identical subgraphs with unitary symmetry described by complex conjugate Hamiltonians, that are coupled by a pair of bonds. The results are compared to random-matrix theory (RMT) predictions obtained on the basis of the Heidelberg approach for the $S$ matrix of open quantum-chaotic systems, employing random matrices from the Gaussian symplectic ensemble. We demonstrate that deviations observed in the distributions of the off-diagonal $S$-matrix elements may be attributed to the fact that the subgraphs are not fully connected, and propose a RMT model, which takes this into account and indeed confirms the experimental results.
\end{abstract}

\maketitle
\date{}
\section{Introduction\label{Intro}}
The field of stochastic quantum scattering emerged in nuclear physics~\cite{Wigner1951,Porter1965}. Indeed, Ericson demonstrated~\cite{Ericson1960,Ericson1963,Ericson1966}, that the cross--sections obtained from nuclear reactions versus the excitation energy fluctuate randomly around their mean, when the resonances strongly overlap. The corresponding energy range is generally referred to as 'Ericson regime'. Chaotic quantum scattering occurs, when waves enter a system with chaotic dynamics through a scattering channel and exit after interacting with the internal states for a sufficiently long time through the same or another one. The theory of quantum-chaotic scattering has been largely developed based on a scattering formalism~\cite{Mahaux1969} introduced for compound-nucleus reactions with in the field of random-matrix theory (RMT). In particular, RMT provides analytical expressions for correlation functions of elements of the scattering ($S$) matrix~\cite{Mahaux1969} that have been generalized to systems with partially or completely violated time-reversal (\T) violation~\cite{Fyodorov2005,Savin2006,Dietz2009}, which are also applicable in the regimes of isolated and weakly overlapping resonances. In the Ericson regime the scattering formalism has been used to discover signatures of \T violation in compound-nucleus reactions~\cite{Witsch1967,Blanke1983}. In fact, it was applied to numerous types of open systems -- mainly in the regime of Ericson fluctuations --, e.g., in nuclear systems~\cite{Brink1963,Brentano1964,Porter1965,Harney1990,Main1992,Feshbach1993,Guhr1998,Weidenmueller2009,Mitchell2010,Frisch2014,Kawano2015}, atomic and molecular systems~\cite{Lombardi1993,Dupret1995,Gremaud1993,Schinke1995,Reid1996,Stania2005,Madronero2005,Mayle2013}, flat microwave billiards modeling quantum billiards~\cite{Doron1990,Stoeckmann1990,Sridhar1994,Alt1995,Sirko1997,Hul2005,Hemmady2006,Dietz2008,Dietz2009a,Dietz2010}, microwave networks as models of quantum graphs~\cite{Hul2004,Kuhl2008,Lawniczak2008,Rehemanjiang2016,Lu2020,Lawniczak2020,Chen2021}, also other wave-dynamical systems~\cite{Ellegaard1995,Rosny2005,Gros2014,Weaver1989}, and is relevant for transport theory~\cite{Bergman1984,Zirnbauer1992,Mirlin1994,Folk1996,Beenakker1997,Alhassid2000,Lee1987,Yeh2012,Bringi2001}.

For quantum-chaotic scattering systems with symplectic symmetry up to now only distributions of the diagonal elements of the $S$ matrix could be determined experimentally with networks of microwave coaxial cables~\cite{Lawniczak2023}. We report on the experimental analysis of the off-diagonal elements of the $S$ matrix of such systems employing again microwave networks. These are described by the telegraph equation. In a certain microwave-frequency range it is mathematically equivalent to that of quantum graphs of corresponding geometry~\cite{Hul2004,Lawniczak2010,Lawniczak2019b}, which are composed of vertices connected by bonds on which they are governed by the one-dimensional Schr\"odinger equation. 

Quantum graphs consisting of bonds of incommensurable lengths provide a suitable testbed for the theoretical study of closed~\cite{Kottos1997,Gnutzmann2004} and open~\cite{Kottos2000,Pluhar2014} quantum-chaotic systems. In distinction to quantum billiards and their microwave analogue all Wigner-Dyson symmetry classes of Dyson's threefold way have been realized with quantum graphs and microwave networks. According to the Bohigas-Giannoni-Schmit conjecture~\cite{Berry1979,Casati1980,Bohigas1984} the spectral properties of typical quantum systems with a chaotic classical dynamics are described by the Gaussian ensemble of random matrices with corresponding universality class~\cite{Mehta2004,Haake2018}. Assuming that there are no additional symmetries~\cite{Dyson1962,Leyvraz1996,Haake2018}, it is either the orthogonal one (GOE) for integer-spin time-reversal (\T) invariant systems~\cite{Hul2004,Lawniczak2008,Hul2012,Lawniczak2016,Dietz2017b,Lawniczak2019}, the unitary one (GUE) when \T invariance is violated~\cite{Hul2004,Lawniczak2010,Lawniczak2019b,Bialous2016,Lawniczak2020,Che2022} or the symplectic one (GSE) for \T invariant half-integer spin systems~\cite{Scharf1988,Dietz1990,Rehemanjiang2016,Lu2020,Che2021,Ramirez2022}.

Quantum graphs were introduced by Linus Pauling~\cite{Pauling1936} as models for organic molecules. Due to their simplicity they are employed as basic mathematical objects~\cite{Imry1996,Pakonski2001,Texier2001,Kuchment2004,Gnutzmann2006,Kowal2009,Berkolaiko2013,Yusupov2019}, also for physical networks in the limit where the lengths of the wires are much larger than their widths~\cite{Kottos1997,Hul2004} and have been used to simulate a large variety of systems, e.g., quantum circuits~\cite{Jooya2016,Namarvar2016}, waveguide systems~\cite{Mittra1971,Post2012,Exner2015,Gnutzmann2022,Zhang2022,Ma2022} and realizations of high-dimensional multipartite quantum states~\cite{Krenn2017}. Following the procedure proposed in~\cite{Rehemanjiang2016,Rehemanjiang2018}, we construct graphs with symplectic symmetry using two geometrically identical subgraphs that are coupled by a pair of bonds. It has been employed in~\cite{Lu2020,Che2021} and in~\cite{Lawniczak2023} to investigate spectral properties and the distributions of the diagonal elements of the $S$ matrix, respectively, of such systems. To obtain information on the off-diagonal $S$-matrix elements of open quantum-chaotic systems with symplectic symmetry, we added scattering channels to the usual construction scheme. 

We compare the experimental distributions to RMT results, which were obtained based on the scattering formalism for compound-nuclear reactions~\cite{Mahaux1969}. The form of the associated $S$ matrix is identical to that of microwave resonators~\cite{Albeverio1996} and quantum graphs and microwave networks~\cite{Kottos1999}. Indeed, numerous experiments~\cite{Bluemel1990,Mendez-Sanchez2003,Schaefer2003,Kuhl2005,Hul2005,Dietz2008,Dietz2009a,Dietz2010,Dietz2010a,Dietz2011a} were performed to model universal properties of the $S$ matrix for compound-nucleus reactions and, generally, for quantum-chaotic scattering processes with preserved or partially violated \T invariance. To be more precise, analytical results~\cite{Verbaarschot1985,Pluhar1995} were derived employing the supersymmetry and RMT approach and validated in Refs.~\cite{Dietz2008,Dietz2009a,Dietz2010} for correlation functions and in Refs.~\cite{Fyodorov2004,Dietz2010} and~\cite{Kumar2013,Kumar2017} for distributions of the diagonal and off-diagonal $S$-matrix elements, respectively. The objective of a still ongoing work is to derive analytical expressions for the distribution of the off-diagonal elements of the $S$ matrix for the symplectic case. In this work we present experimental results and outline the difficulties that are encountered in their analytical description. 

In~\refsec{SecExp} we review the experimental realization of quantum graphs with symplectic symmetry, referred to as symplectic graphs in the following. Then, in~\refsec{SecNum} we outline how we compute the eigenstates and $S$ matrix of the corresponding quantum graphs~\cite{Kottos1999,Kottos2003}. In~\refsec{SecDistr} we present the experimental and numerical results for the distributions of the off-diagonal elements of the $S$ matrix of these symplectic graphs. We demonstrate that they do not agree with RMT predictions deduced from the Heidelberg approach, which employs Hermitean random matrices from the GSE. This is attributed to the fact, that the two subgraphs that form the symplectic graph are not fully connected. To confirm this assumption, we in addition performed simulations with random $S$ matrices constructed on the basis of the same scattering formalism, however with modified Hermitean random matrices, that resemble the symplectic graph. Finally, we summarize our findings in~\refsec{SecConcl}. 

\section{Experimental realization of graphs with symplectic symmetry\label{SecExp}}
In the first experiments with microwave networks modeling quantum graphs with symplectic symmetry, put forward in Ref.~\cite{Rehemanjiang2016} and in Refs.~\cite{Lu2020,Che2021}, spectral properties were investigated and found to be well described by the GSE. The symplectic graphs are designed such that the corresponding Hamiltonian~\cite{Rehemanjiang2016,Lu2020} attains in the basis $\mathcal{W}=\{\vert 1\uparrow\rangle,\vert 2\uparrow\rangle,\dots,\vert N\uparrow\rangle,\vert 1\downarrow\rangle,\vert 2\downarrow\rangle,\dots,\vert N\downarrow\rangle\}$ the form
\be
H=\begin{pmatrix}
H_0 &V\\ -V^\ast &H_0^\ast\end{pmatrix}, \quad  H_0=H_0^\dagger, \quad V=-V^T.\label{Ham}
\ee
The matrix $V$ couples the matrix $H_0$ of the spin-up system, which is drawn from the GUE, with its complex conjugate which describes the spin-down system. A Hamiltonian with this structure exhibits symplectic symmetry, $Y H^{\text{T}} Y^{\text{T}} = H$ with $Y=-i\tau^{(2)} \otimes \mathds{1}_N$, where $\tau^{(2)}$ denotes the second Pauli matrix. It belongs to the GSE only for nonzero matrices $V$~\cite{Dyson1962}, because only then it complies with the condition that in a system with symplectic symmetry spin-rotation symmetry is broken~\cite{Dyson1962}. 

\begin{figure}[htpb]
\includegraphics[width=1.0\linewidth]{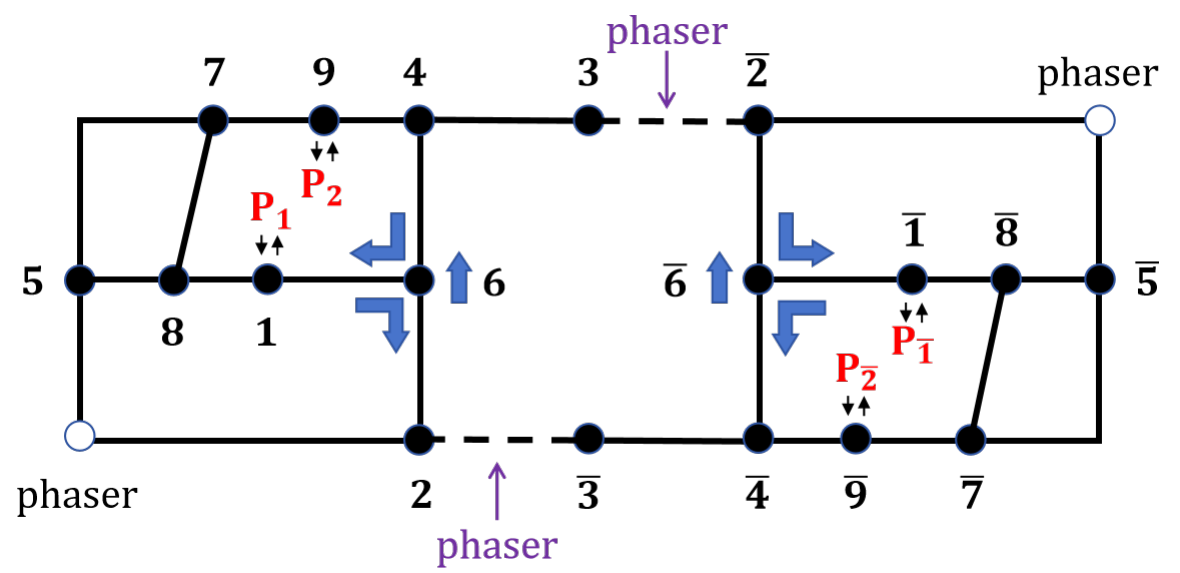}
\caption{Schematic view of the microwave network with symplectic symmetry constructed from coaxial cables that are connected at T joints. The blue arrows indicate the effect of the T-shaped circulators. The coaxial cables connecting subgraphs marked by $(3,\bar 2)$ and $(2,\bar 3)$ are connected to phase shifters, indicated by dashed-line bonds. Additional phase shifters are connected to the bonds marked by circles to change the lengths of bonds. The four antennas are attached at ports $P_1, P_2, P_{\bar 1}$ and $ P_{\bar 2}$.}\label{Fig1}
\end{figure}

A scheme of the symplectic graph used for the numerical and experimental studies is depicted in~\reffig{Fig1}. It consists of two geometrically identical GUE subgraphs belonging to the unitary universality class and modeling $H_0$ and $H_0^\ast$ in~\refeq{Ham}, with the vertices marked by indices $i$ and $\bar i$, corresponding to spin orientations $\uparrow$ and $\downarrow$, respectively. Each of the subgraphs consists of $\mathcal{V}=9$ vertices and the whole graph has $\mathcal{B}=22$ bonds connecting them. The geometry of the graph is defined by the connectivity matrix $C$, which has vanishing diagonal elements $C_{\tilde i\tilde i}=0$ and its off-diagonal elements are $C_{\tilde i\tilde j}=1$ if vertices $\tilde i$ and $\tilde j$ are connected and $C_{\tilde i\tilde j}=0$ otherwise. Corresponding bond lengths coincide, $L_{ij}=L_{\bar i \bar j}$. 

The experimental graph is a microwave network, which is constructed from coaxial microwave cables that are connected by T joints, corresponding to the bonds and vertices, respectively. Time-reversal invariance violation is induced by T-shaped circulators of opposite orientation introduced at corresponding vertices, that lead to a directionality as illustrated by the blue arrows, and may be described by an additional phase $\pi/2$, respectively $-\pi/2$ in the waves. The subgraphs are connected at two vertex pairs marked by $i_0=2$, $\bar j_0=\bar 3$ and $j_0=3$, $\bar i_0 =\bar 2$, respectively, that have same lengths $L_{i_0 \bar j_0}=L_{j_0 \bar i_0}$. The requirement $V=-V^T$, that is, $V_{i_0 \bar j_0}=-V_{j_0 \bar i_0}$, is realized with an additional phase of $\pi$ on one of the bonds, e.g., on that connecting $i_0$ and $\bar j_0$. Different realizations of the symplectic graph were obtained by increasing the lengths of two corresponding bonds marked by circles in~\reffig{Fig1} with phase shifters by the same amount. The absorption strength was changed by introducing 1~dB to 30~dB attenuators. 

The experiments were conducted with four instead of two~\cite{Lawniczak2023} antennas attached to four ports $P_1, P_2, P_{\bar 1}$ and $ P_{\bar 2}$, respectively. For the measurements of the $S$-matrix elements the antennas were connected to an Agilent N5227A vector network analyzer (VNS) via SUCOFLEX126EA/11PC35/1PC35 coaxial-cables emitting microwaves into the resonator via one antenna $a$ and receiving it at an other one, $b$. The VNA measures relative phases and ratios of the outgoing and incoming microwaves as function of their frequency $f$, yielding the complex $S$-matrix elements $S_{ba}(f)$.

\begin{figure}[htpb]
\begin{center}
\includegraphics[width=1.\linewidth]{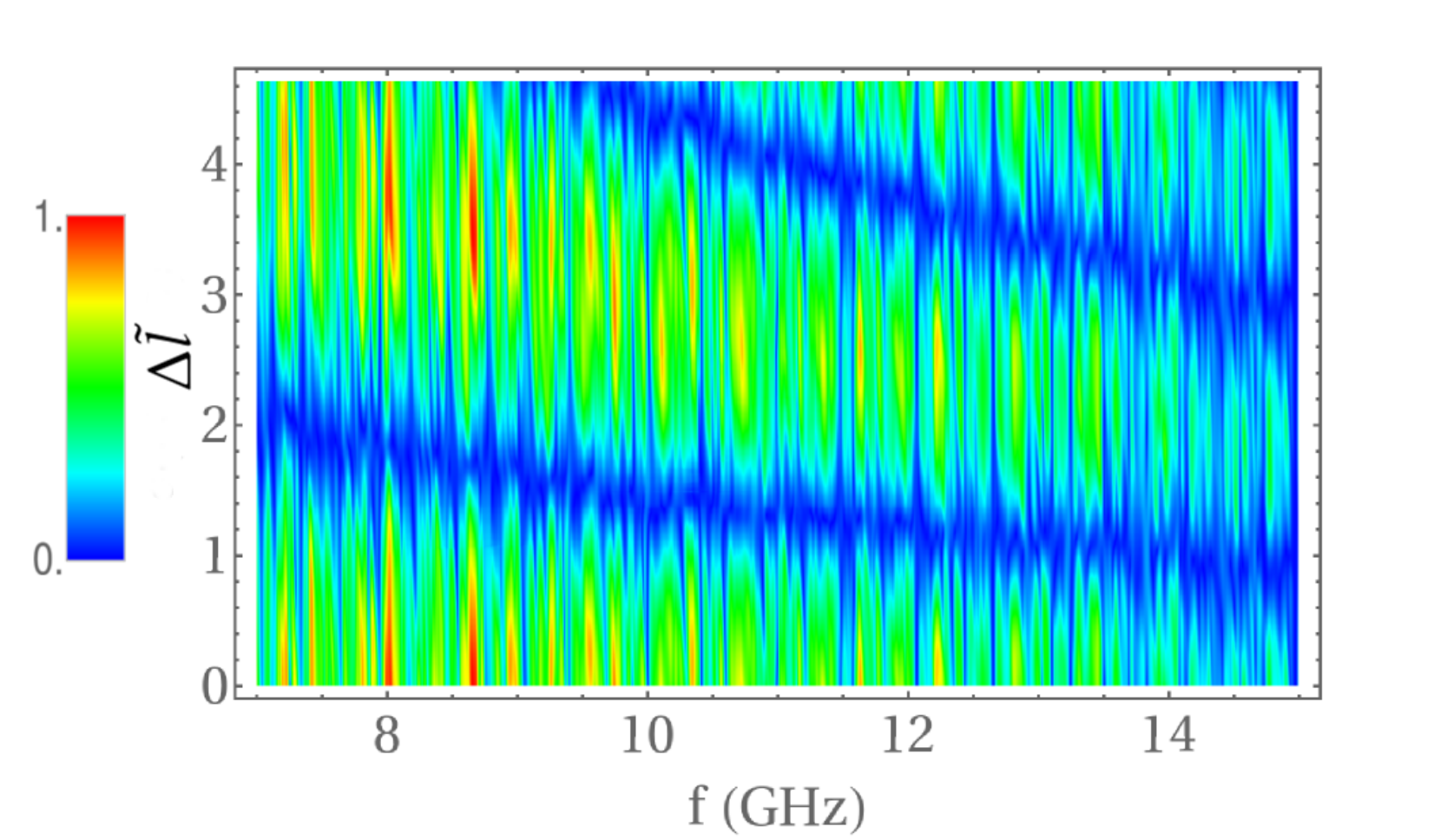}
\caption{Density plot of the transmission amplitude $\vert S_{1\bar 1}(f)\vert$ versus microwave frequency $f$ for fixed relative length differences $\Delta\tilde l$ between the bonds connecting vertices $2$ and $\bar 3$, respectively, $3$ and $\bar 2$.}\label{Fig3}
\end{center}
  \end{figure}

The phase difference of $\pi$ on the two bonds connecting the two GUE graphs was generated by a phase shifter which, actually, changes the length of the coaxial cable by some increment $\Delta \tilde l$ and thus induces a change of the phase accumulated by microwaves passing through it according to the relation
\begin{equation}
\Delta\varphi=k\Delta\tilde l=\frac{2\pi f}{c}\Delta\tilde l \ .
\label{DeltaL}
\end{equation}
Hence, the phase shift $\Delta\varphi$ depends on the microwave frequency $f$ or wavenumber $k=2\pi f/c$. Therefore the experiments were performed for 350 values of $\Delta\tilde l$ and a phase difference of $\pi$ was identified as vanishing transmission --- to be more precise, in the experiments as minimal transmission --- between $P_1$ and $P_{\bar 1}$, or $P_2$ and $P_{\bar 2}$, respectively. In~\reffig{Fig3} all 350 transmission spectra between $P_1$ and $P_{\bar 1}$ are plotted versus frequency $f$. Here, blue color corresponds to vanishing transmission and red color to maximal transmission. The phase differences of $\pi$, and $3\pi$ correspond to the blue stripes. Actually, the phase varies continuously, corresponding to a change from GSE to GOE with $\Delta\varphi=0, 2\pi$ and GUE in between. In this transition region the condition for symplectic symmetry, $V=-V^T$, does not hold and $-V^\ast$ in~\refeq{Ham} needs to be replaced by $V^\dagger$.

\section{Numerical analysis of the corresponding quantum graphs\label{SecNum}}
We also performed numerical simulations of quantum graphs for which the wave-function component $\psi_{\tilde i\tilde j}(x)$ on the bond connecting vertices $\tilde i$ and $\tilde j$ is a solution of the one--dimensional Schr\"odinger equation 
\be
\left(-i\frac{d}{dx}-A_{\tilde i\tilde j}\right)^2\psi_{\tilde i\tilde j}(x)+k^2\psi_{\tilde i\tilde j}(x)=0. 
\label{Schr}
\ee
Here, the coordinate $x$ varies along the bond from $x=0$ at vertex $\tilde i$ to $x=L_{\tilde i\tilde j}$ at vertex $\tilde j$, and $A=-A^T$ denotes the magnetic vector potential which induces violation of time--reversal invariance. Furthermore, $A_{ij}=-A_{\bar i \bar j}$ on corresponding bonds, since one of the GUE graphs should be the complex conjugate of the other one. On the bonds that couple the GUE graphs we set $A_{\tilde i\tilde j}=0$. The wave-function components are subject to boundary conditions imposed at the vertices that ensure continuity and conservation of the current~\cite{Kottos1999}. The microwave networks model quantum graphs with Neumann boundary conditions at the vertices~\cite{Hul2004}. They constitute a special case of $\delta$-type boundary conditions~\cite{Kottos1999,Kostrykin1999,Texier2001,Kuchment2004}. It should be noted that the wave equation for microwave networks is mathematically equivalent to that of the corresponding quantum graph only below the cutoff frequency for the first transverse electric mode, where only the fundamental transverse electromagnetic mode can propagate between the conductors
\cite{Jones1964,Savytskyy2001}.
	
The eigenwavenumbers of closed quantum graphs with these boundary conditions are determined by solving the equation~\cite{Kottos1999}, $\det H(k)=0$, with
\begin{equation}
H_{\tilde i\tilde j}(k)=\left\{{\begin{array}{cc}\displaystyle
    -\sum_{\tilde m\ne\tilde i}\cos\left(kL_{\tilde i\tilde m}\right)
        \frac{C_{\tilde i\tilde m}}{\sin\left(kL_{\tilde i\tilde m}\right)}\ , & \tilde i=\tilde j\\
	\displaystyle e^{-i\hat A_{\tilde i\tilde j}L_{\tilde i\tilde j}-i\Phi_{\tilde i\tilde j}}
	\frac{C_{\tilde i\tilde j}}{\sin(kL_{\tilde i\tilde j})}\ , & \tilde i\ne\tilde j\\
        \end{array}} \right. \ ,
\label{QuantE}
\end{equation}
where $\Phi_{\tilde i\tilde j}=\pi$ for $(\tilde i=i_0, \tilde j=\bar j_0$) and zero otherwise. For the magnetic vector potential we chose $\vert A_{\tilde i\tilde j}\vert =\frac{\pi}{2}$ on the bonds connecting vertices $i=1,2,4$ with vertex $j=6$ and $\bar i=\bar 1,\bar 2,\bar 4$ with vertex $\bar j=\bar 6$, respectively, such that $A_{ij}=-A_{\bar i \bar j}$ (see~\reffig{Fig1}). The components of the associated eigenvectors yield the values of the wave functions at the vertices and thus the eigenfunctions~\cite{Kottos1999}. 

\begin{figure}[htpb]
\includegraphics[width=1.0\linewidth]{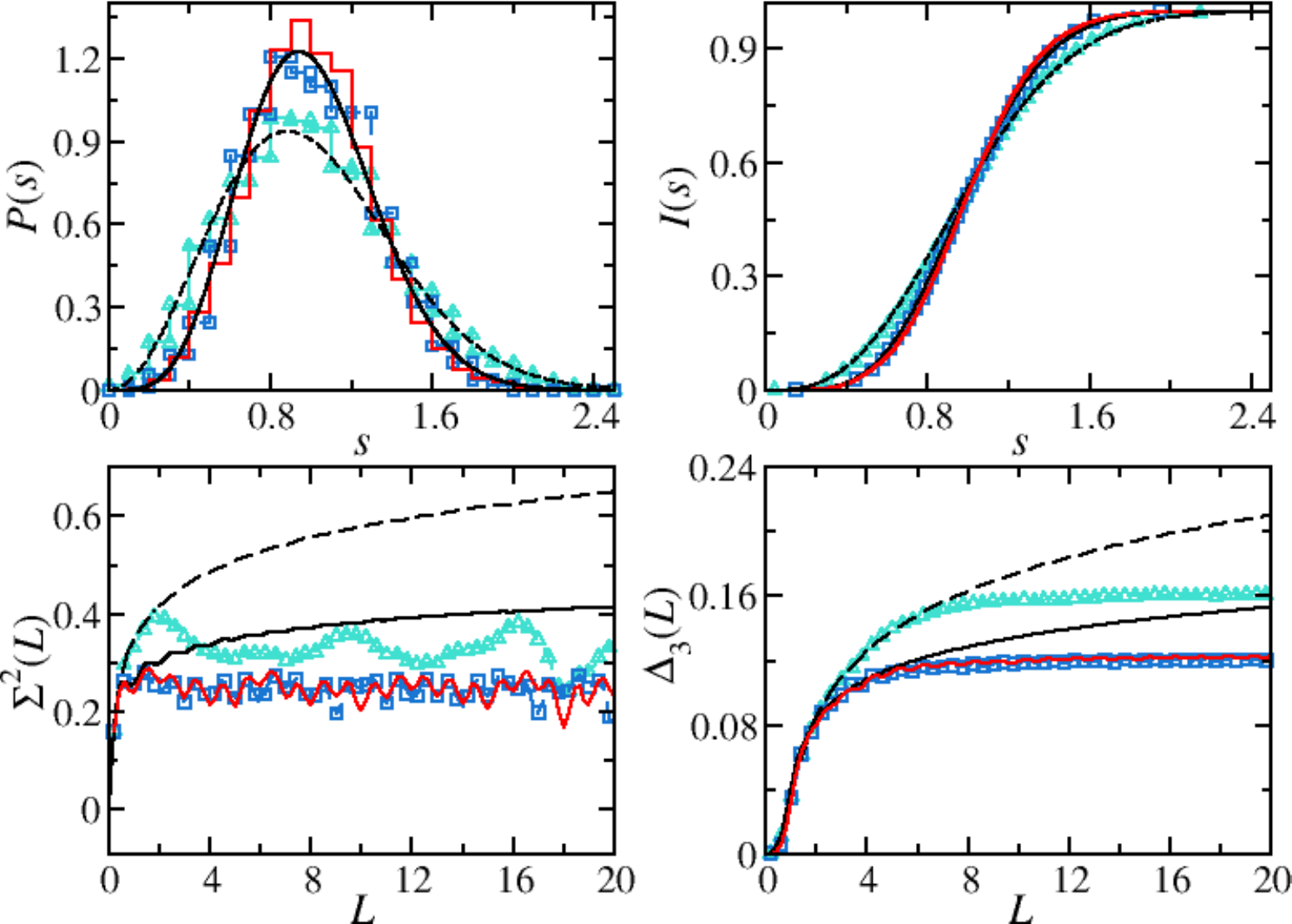}
	\caption{Nearest-neighbor spacing distribution $P(s)$, integrated nearest-neighbor spacing distribution $I(s)$, number variance $\Sigma^2(L)$ and rigidity $\Delta_3(L)$, with (red lines) and without (turquoise lines with triangles) magnetic field. Black solid and dashed lines show the GSE and GUE curves, respectively. Blue lines with squares show the results for the symplectic graph with four coupling bonds instead of only two.}\label{Fig2}
\end{figure}

We computed 8000 eigenvalues $k_j$ employing~\refeq{QuantE}. Before analyzing fluctuations, the spectra need to be unfolded to mean spacing unity. This is trivial for quantum graphs, because the spectral density is constant, $\rho(k)=\mathcal{L}/\pi$ where $\mathcal{L}$ denotes the total length of the graph. In~\reffig{Fig2} several statistical
measures for the spectral properties are shown for the graph exhibited in~\reffig{Fig1}, also for the case with no magnetic field, $A_{\tilde i\tilde j}=0\, \forall \tilde i\tilde j$. In both cases the Hamiltonian takes the form \refeq{Ham}, however, $H_0=H_0^\ast$ in the latter case. In addition, we performed numerical simulations for the GUE graph~\reffig{Fig1} with four coupling bonds instead of only two~\cite{Rehemanjiang2018}. To obtain it, we removed the bonds between vertices $7$ and $8$ and between $\bar 7$ and $\bar 8$, respectively, and connected the vertex $7$ with the vertex $\bar 8$ and $8$ with $\bar 7$. The spectral properties of both graphs agree well, for the short-range correlations also with the GSE curves, that is, the level repulsion is quartic, thus corroborating that the graph exhibits symplectic symmetry. Furthermore, this indicates high spectral rigidity which is reflected in the logarithmic dependence of the number variance $\Sigma^2(L)$ and the rigidity $\Delta_3(L)$ on the interval length $L$, that is, number of mean spacings, in contrast to a linear one exhibited by uncorrelated levels~\cite{Berry1985,Heusler2007}. These are shown in the lower part of~\reffig{Fig2}. Clear deviations of the long-range correlations from the RMT predictions are observed beyond $L\gtrsim 2.5$ for $\Sigma^2(L)$ and $L\gtrsim 5$ for $\Delta_3(L)$. They are attributed to nonuniversal contributions. Namely, quantum graphs with Neumann boundary conditions comprise eigenfunctions that are localized only on a fraction of the bonds~\cite{Dietz2017b,Lu2020}. 

For the numerical simulations of the $S$ matrix describing the reflection spectra, $S_{aa}$, and transmission spectra, $S_{ab}$, from port $P_a$ at vertex $j^\prime$ to port $P_b$ at vertex $i^\prime$ of the microwave networks with $a,b\in{1,2,\bar 1,\bar 2}$, we used the following one, given in~\cite{Texier2001,Kottos2003,Pluhar2013,Pluhar2013a,Pluhar2014}, 
\begin{equation}
\label{S_Graph}
S_{ba}^\mathcal{B}=\delta_{i^\prime,j^\prime}\rho^{(i^\prime)}+\sum_{i,j}\tau_i^{(i^\prime)}\left[\II-S_B(k)\right]^{-1}_{i^\prime i,jj^\prime}D_{(jj^\prime,jj^\prime)}\tau_j^{(j^\prime)} \ ,
\end{equation}
where $S_B(k)=D(k)U$ is defined in the $2\mathcal{B}$ space of directed bonds with indices $i^\prime i,j j^\prime$ referring to the bonds between vertices $i^\prime ,i$, and $j ,j^\prime$, respectively. The matrix
\begin{equation}
D_{ij,nm}=\delta_{i,n}\delta_{j,m}C_{ij}e^{ikL_{ij}+iA_{ij}L_{ij}+\Phi_{ij}},\label{DMatrix}
\end{equation}
is diagonal in the $2\mathcal{B}$ space and the transition from vertex $m$ via vertex $i$ to vertex $j$ is described by
\begin{equation}
U_{ji,nm}=\delta_{n,i}C_{ji}C_{nm}\Sigma^{(i)}_{jm} \ ,
\end{equation}
where the vertex $S$ matrix at vertex $i$ is defined as $\Sigma^{(i)}_{nm}=\frac{2}{\tilde v_i}-\delta_{n,m}$, $\tau_m^{(i)}$ is nonzero only at vertices $i^\prime $ connected to a lead, $\tau_m^{(i^\prime)}=2/\tilde v_{i^\prime}$, $\rho^{(i^\prime)}=2/\tilde v_{i^\prime}-1$ at vertices $i^\prime$ coupled to leads, and $\rho^{(i)}=1$ on the other vertices $i$. Here, $\tilde v_i$ denotes the total number of bonds and leads connected to vertex $i$. To account for absorption, we added a small imaginary part to the wave number $k$, $k\to k+i\epsilon$. The expression~\refeq{S_Graph} can be brought to a form~\cite{Kottos2003} which is similar to that deduced from the $S$-matrix formalism for compound nucleus reactions~\cite{Mahaux1969,Verbaarschot1985}, 
\begin{align}
S_{ab}(k) &= \delta_{ab}-i 2\pi W_a^\dag G(k) W_b\, ,\\
	G^{-1}(k) &= k\mathds{1}_{2N}-H+i\pi\displaystyle\sum_{c=1}^{2M} W_c W_c^\dag\, ,
\label{Sab}
\end{align}
except that $H(k)$ depends explicitly on $k$.
Here, in the $\mathcal{W}$ basis $a,b,c\in 1,2,\dots,M,\bar 1,\bar 2,\dots \bar M$ for the case of $M$ leads connected to corresponding vertices in the subgraphs, yielding $2M$ scattering channels and thus a $(2M)\times(2M)$ dimensional $S$ matrix. Furthermore, $W_c$ accounts for the interaction between the scattering channels $c$ and the states of $H$. 
We obtained the eigenwavenumbers of the closed graph either with~\refeq{QuantE} or with~\refeq{S_Graph} by solving the secular equation~\cite{Kottos2003}, $\zeta_\mathcal{B}(k)=\det\left[\II-S^\mathcal{B}(k)\right]=0$. Both procedures yield the same result, thus corroborating the validity of the $S$-matrix ansatz.
\begin{figure}[htbp]
  \begin{center}
\includegraphics[width=1.\linewidth]{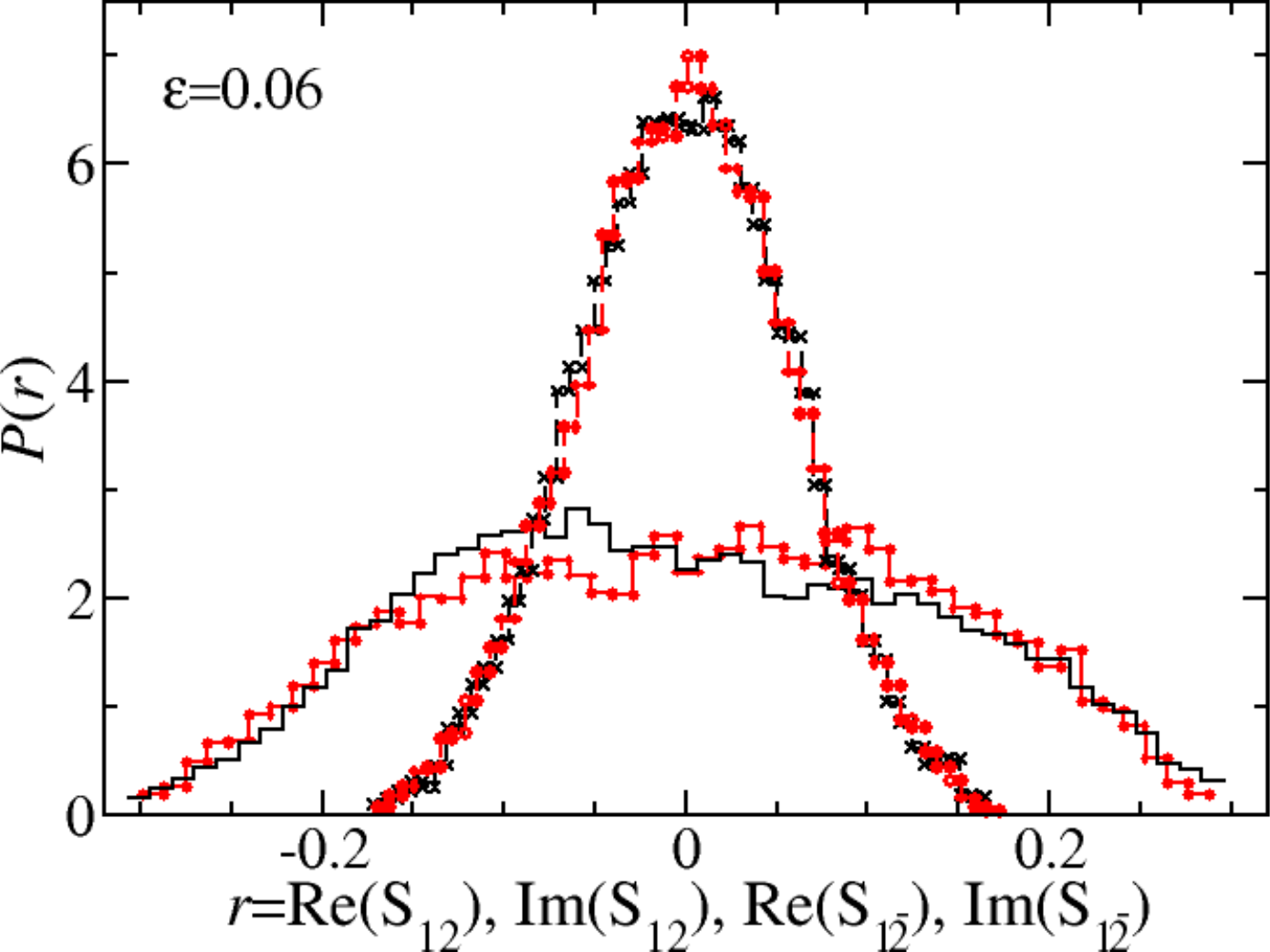}
\includegraphics[width=1.\linewidth]{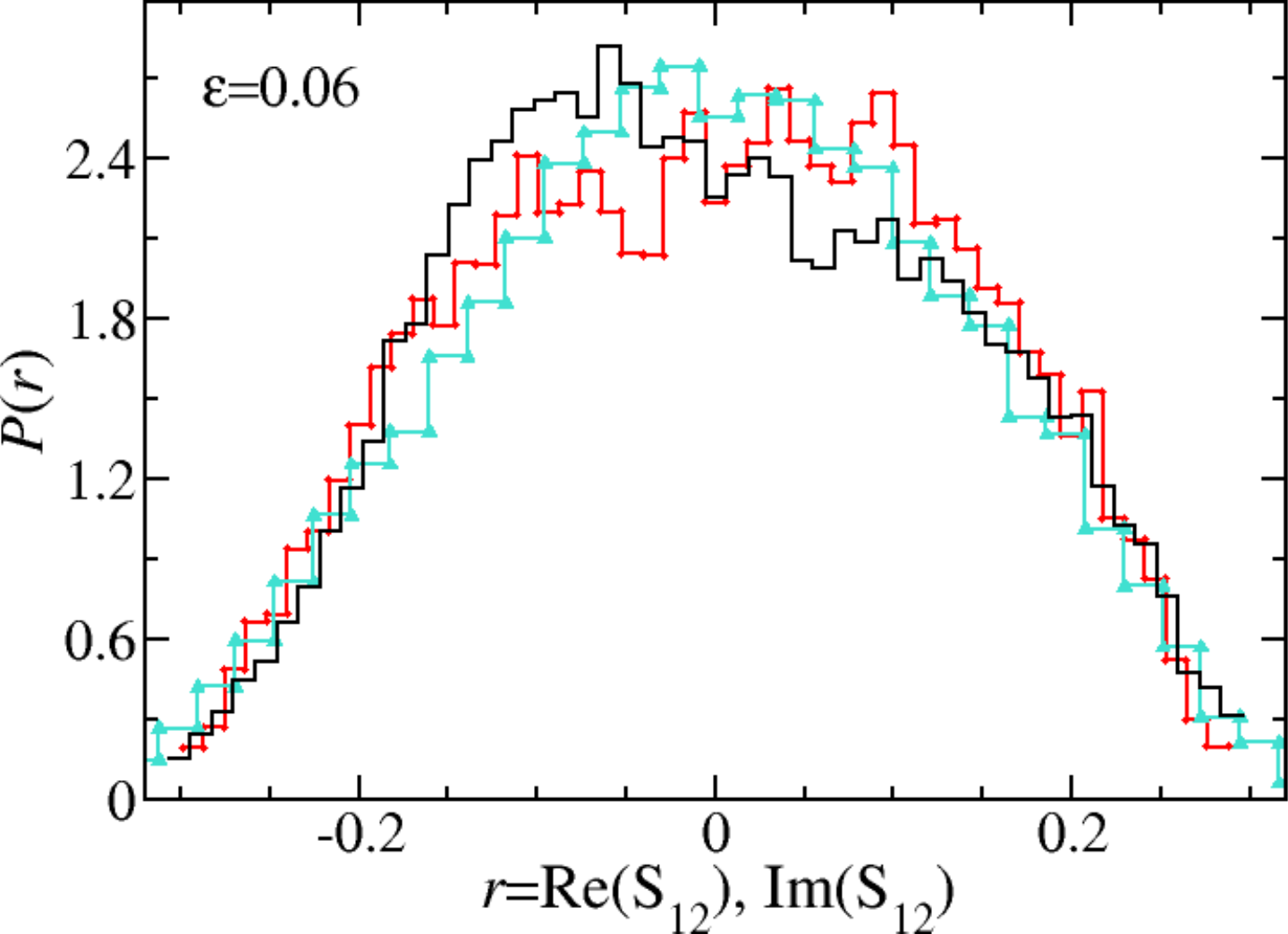}
\includegraphics[width=1.\linewidth]{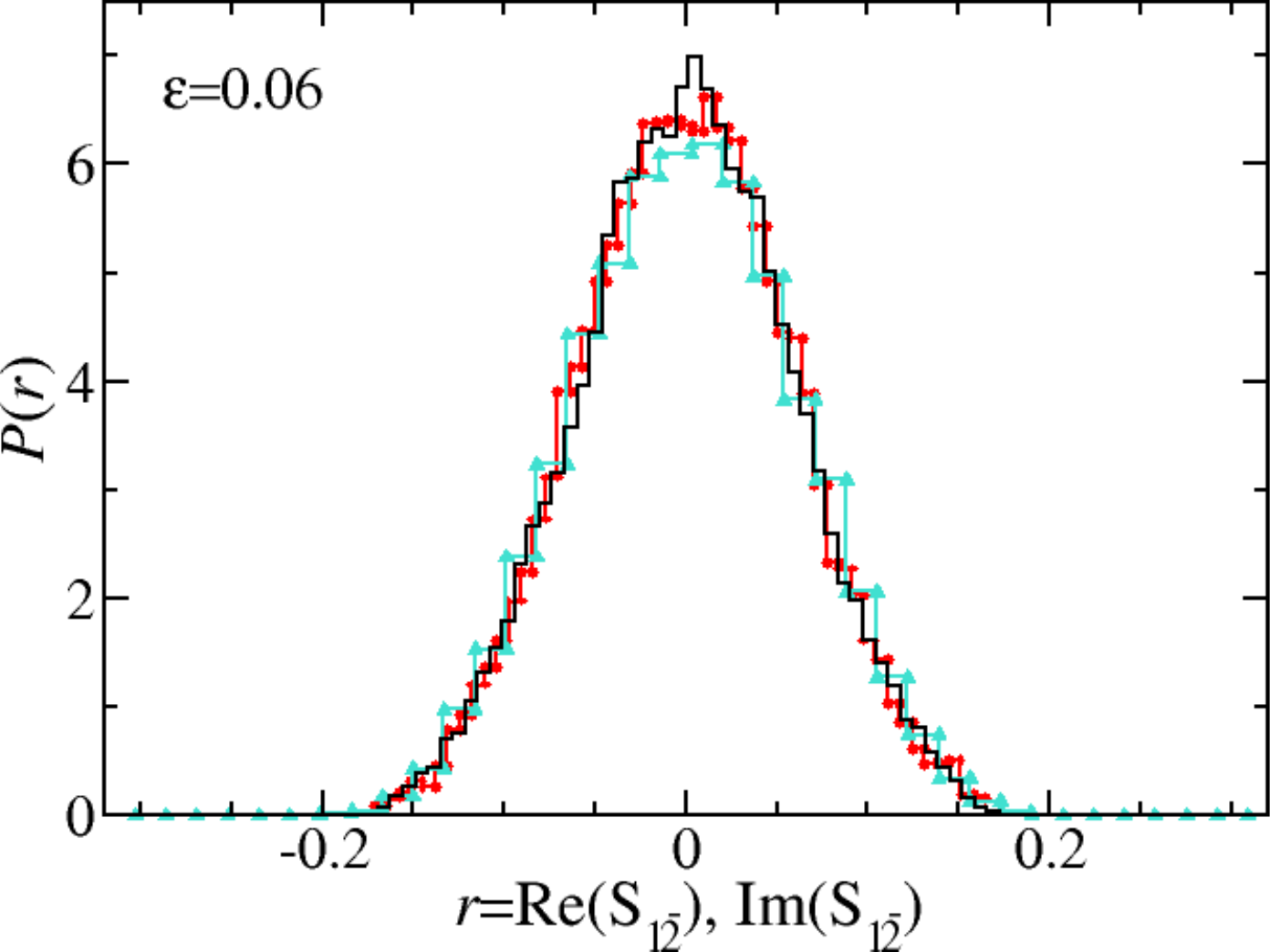}
\caption{Top: Comparison of the experimental distributions of the real and imaginary parts for $S_{12}$ (black line and red solid line with dots) and for $S_{1\bar 2}$ (black dashed line with crosses and red dashed line with dots). Middle: Comparison of the distributions for the real and imaginary parts of $S_{12}$ (black line and red line with dots) with the numerical results (turquoise line with triangles) for the corresponding quantum graph obtained from~\refeq{S_Graph} for the imaginary parts. The numerical distribution for the real part lies on top of the three other distributions and therefore it is not shown. Bottom: Same as top for $S_{1\bar 2}$.
          }\label{Fig4}
\end{center}
\end{figure}
\begin{figure}[htbp]
\begin{center}
\includegraphics[width=0.8\linewidth]{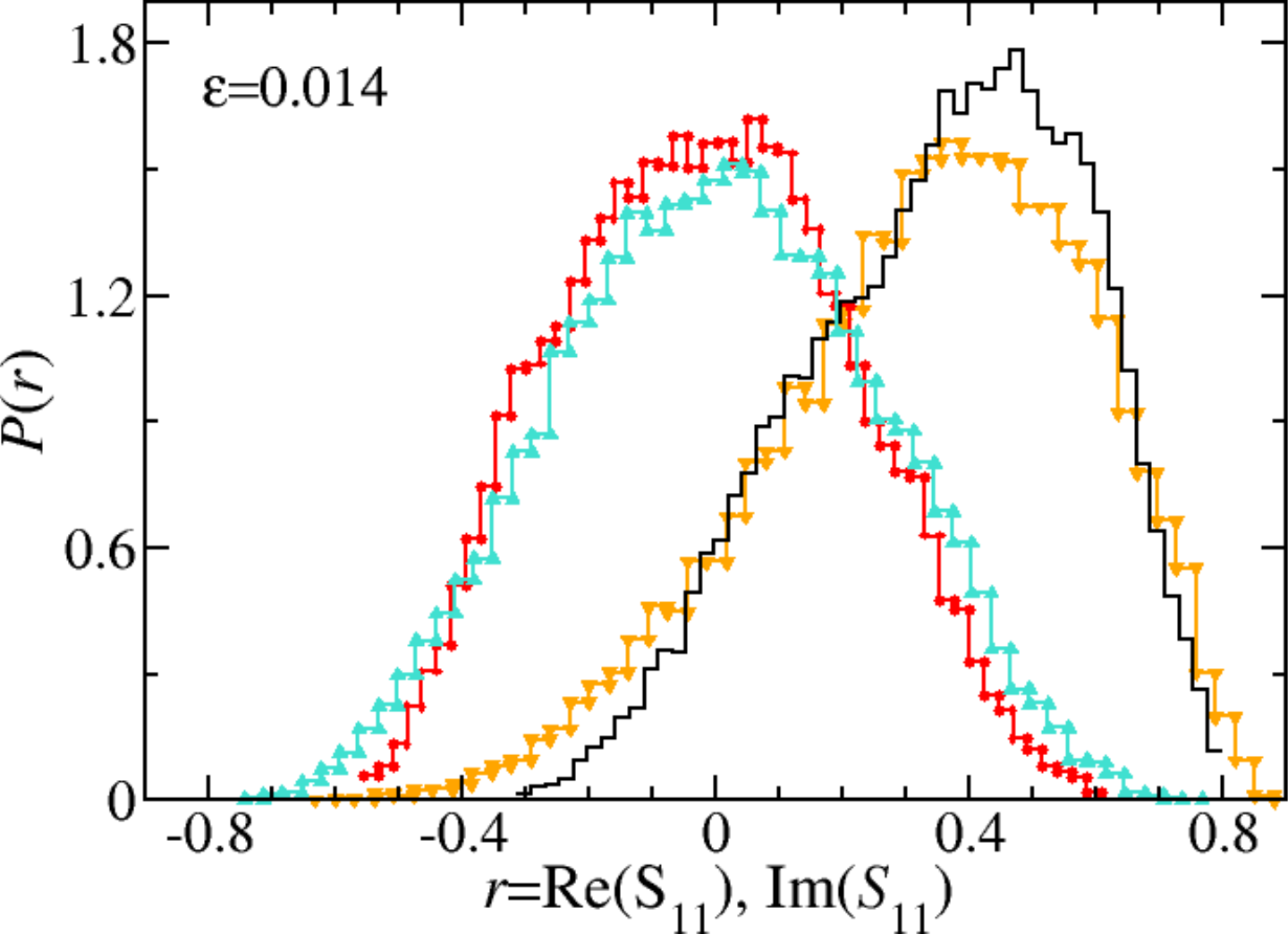}
\includegraphics[width=0.8\linewidth]{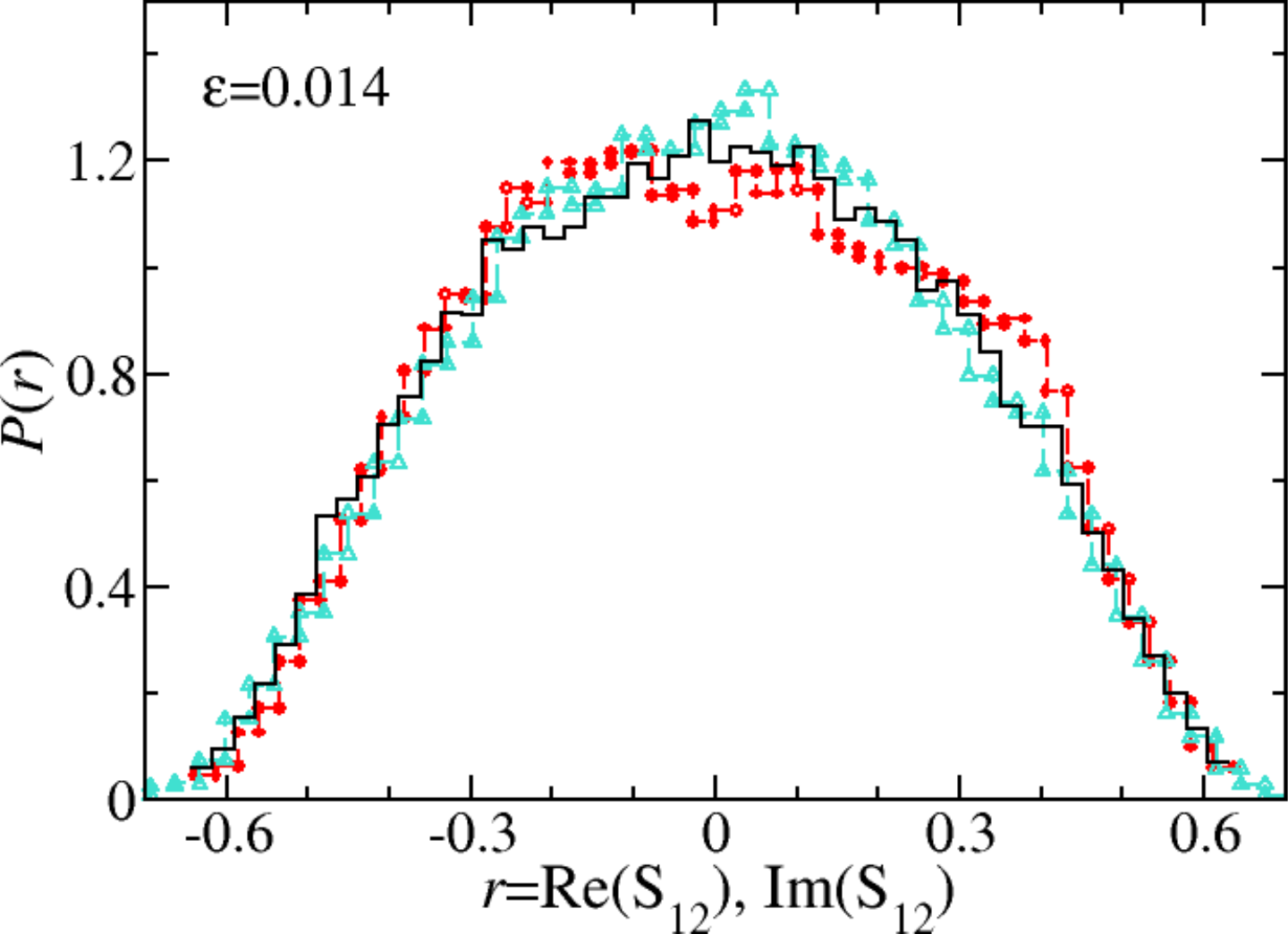}
\caption{Top: Comparison of the experimental distributions of the real and imaginary parts for $S_{11}$ (black line and red line with dots) with the corresponding numerical ones for the corresponding quantum graph (orange line with triangles down and turquoise line with triangles up) obtained from~\refeq{S_Graph}. Bottom: Comparison of the distributions for the real and imaginary parts of $S_{12}$ (black line and red line with dots) with the numerical results (turquoise line with triangles) obtained from~\refeq{S_Graph}. The numerical distribution for the real part is not shown, because it lies on top of the other three curves.
}\label{Fig5}
\end{center}
\end{figure}
\begin{figure}[htbp]
\begin{center}
\includegraphics[width=0.9\linewidth]{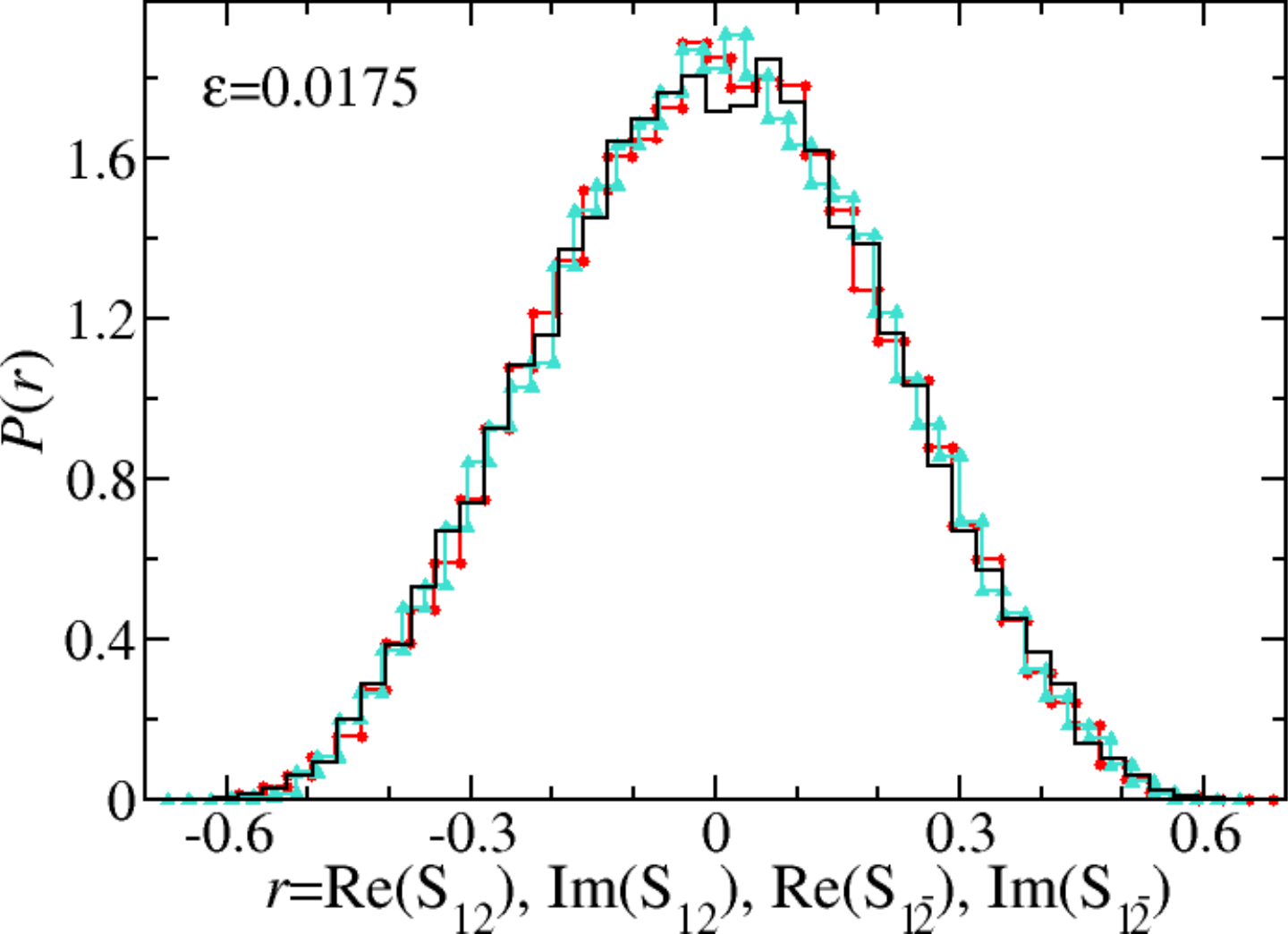}
\caption{Comparison of the distributions of the real part of $S_{12}$ (black line) and $S_{1\bar 2}$ (red line with dots) and the imaginary part of $S_{1\bar 2}$ (turquoise line with triangles) for four coupling cables obtained from the numerical computations for the quantum graph. The results for the distribution of the imaginary part of $S_{12}$ is not shown, because it lies on top of the other curves. Results are shown for absorption $\epsilon =0.0175$.}
\label{Fig6}
\end{center}
\end{figure}
\begin{figure}[htbp]
\begin{center}
\includegraphics[width=0.8\linewidth]{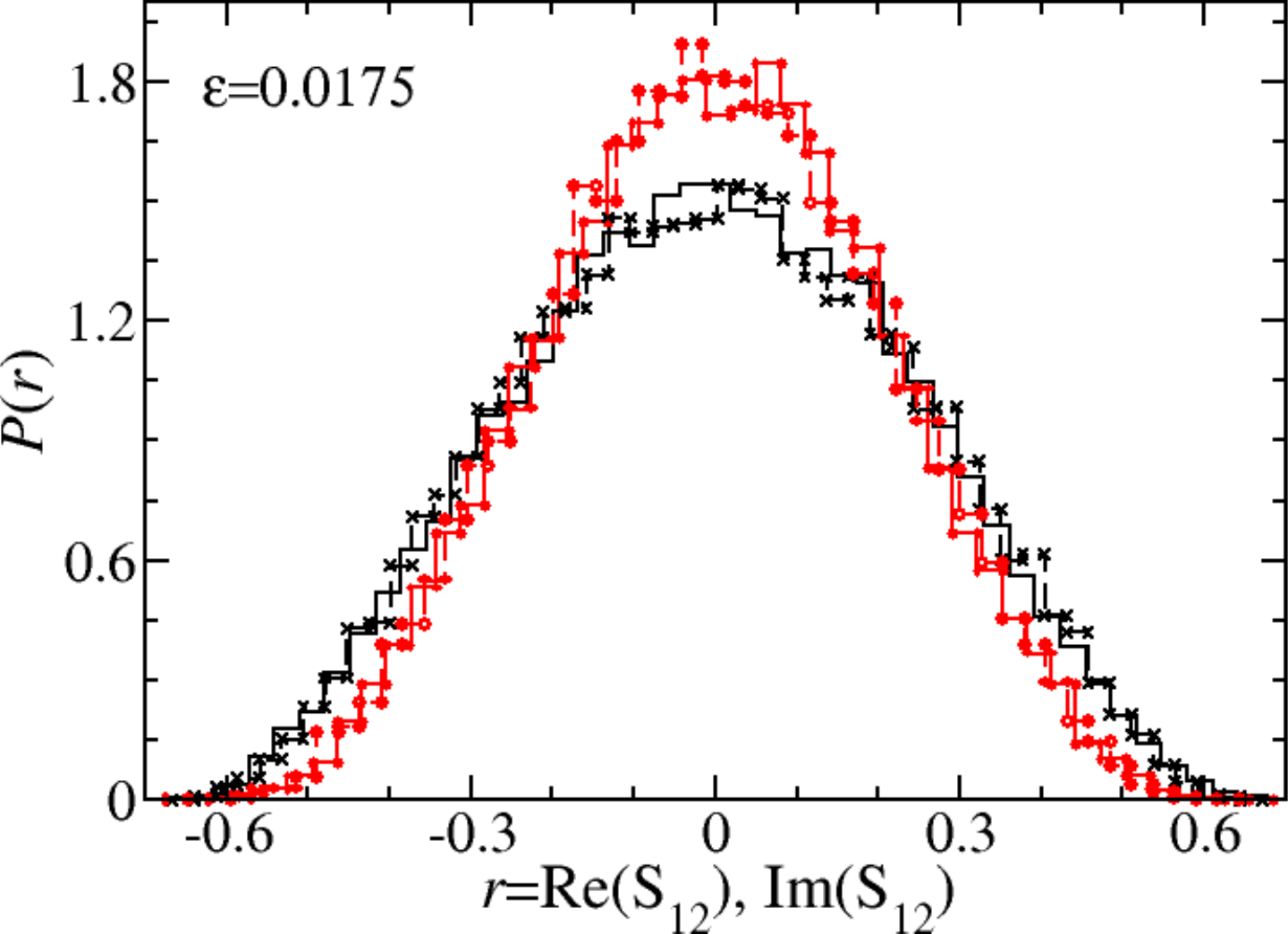}
\includegraphics[width=0.8\linewidth]{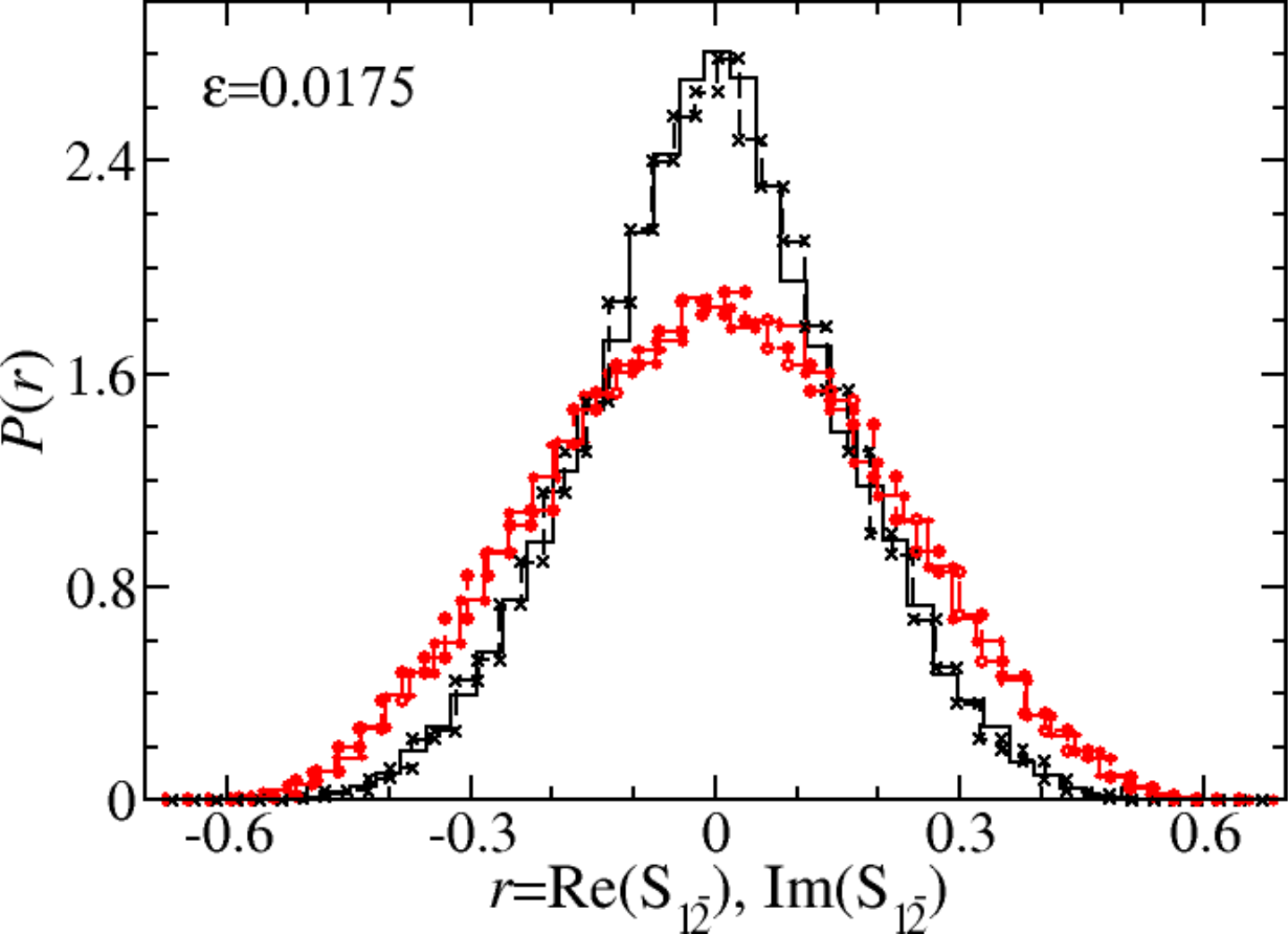}
\includegraphics[width=0.8\linewidth]{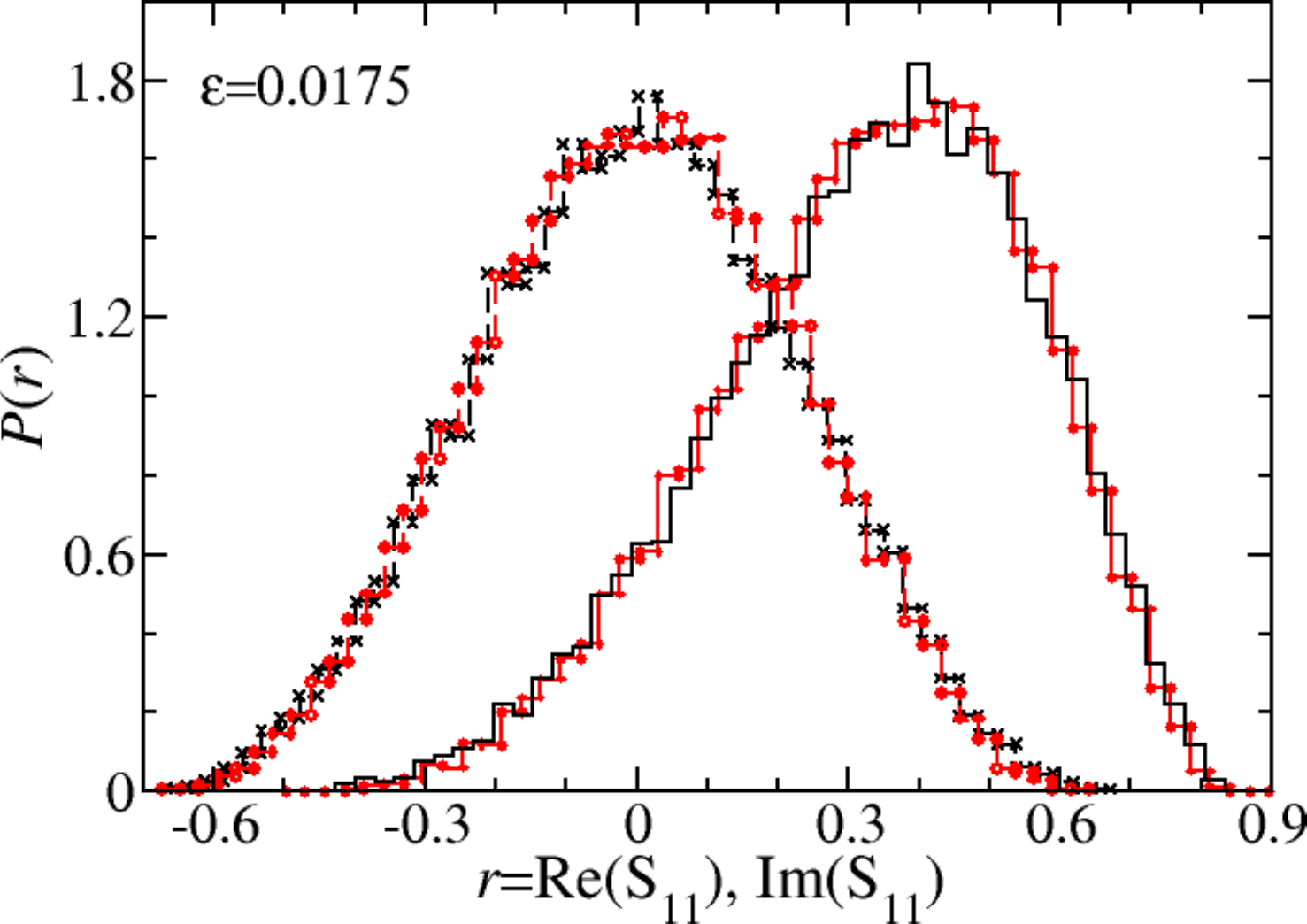}
\caption{
Top: Comparison of the distributions of the real and imaginary parts for $S_{12}$ (solid and dashed lines) with two coupling bonds (black solid line and dashed line with crosses) with those for four coupling bonds (red lines with dots) obtained from the numerical computations for the quantum graph with $\epsilon =0.0175$. Middle: Same as top for $S_{1\bar 2}$. Bottom: Same as top for $S_{11}$.}
\label{Fig7}
  \end{center}
\end{figure}
\begin{figure}[htbp]
\begin{center}
\includegraphics[width=0.9\linewidth]{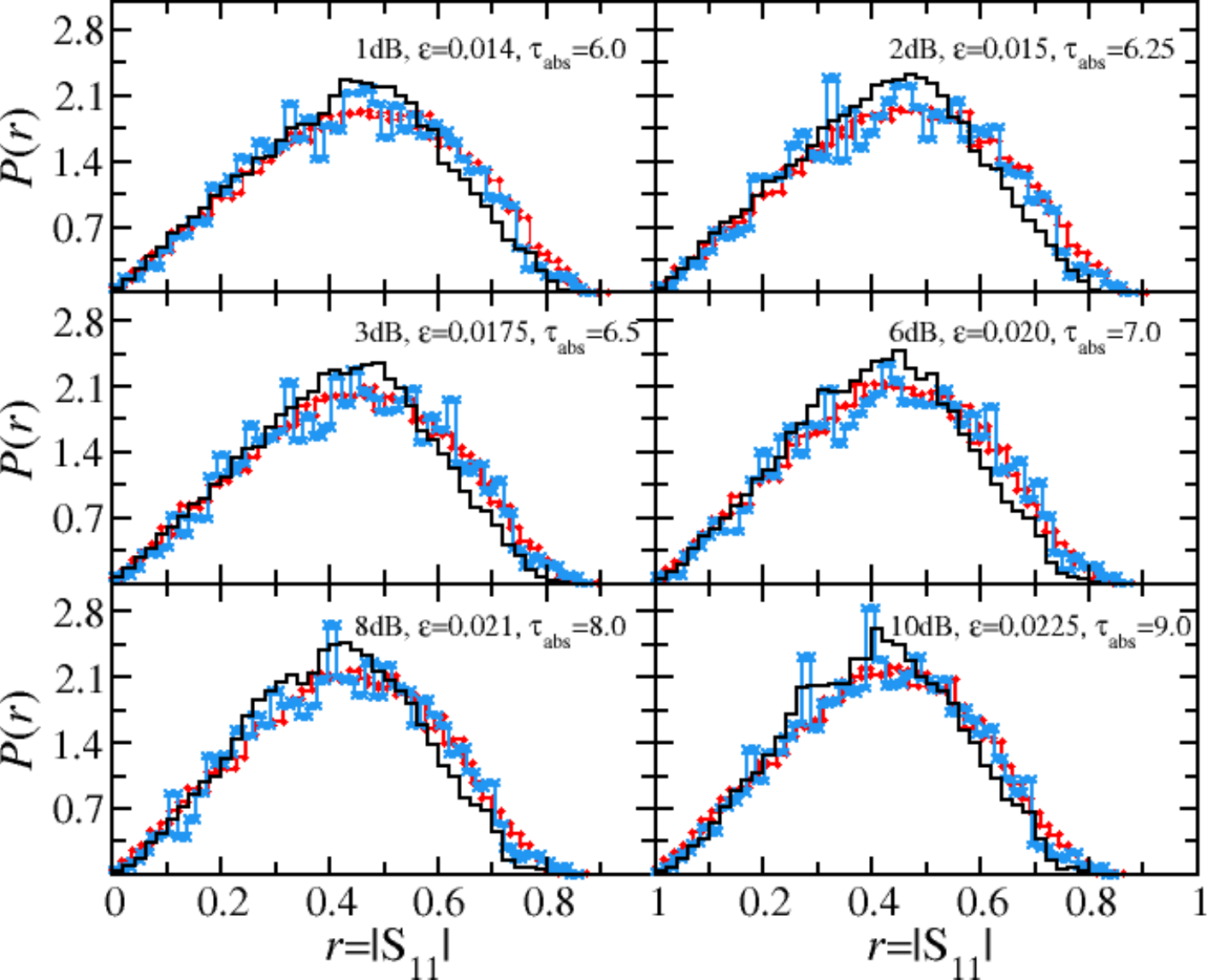}
\includegraphics[width=0.9\linewidth]{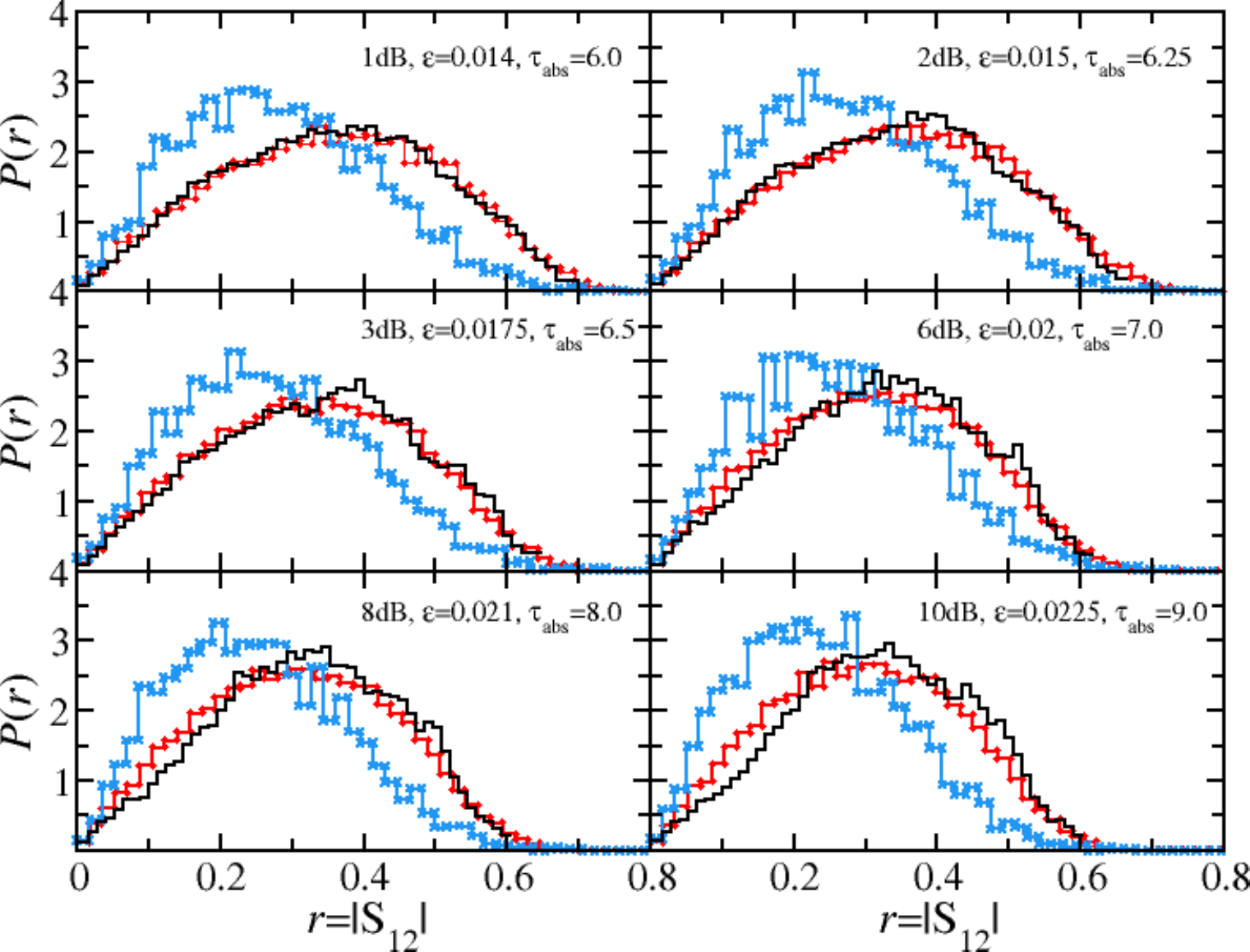}
\caption{Top: Comparison of the experimental (black lines) and numerical (red lines with dots) distributions of the modulus of $S_{11}$ with RMT simulations (blue lines with crosses). The values of the absorption strength induced in the experiments by attenuators, of $\epsilon$ in the quantum graphs and of the absorption parameter $\tau_{abs}$ in the RMT simulations are given in the panels. Bottom: Same as top for $S_{12}$.
          }\label{Fig8}
  \end{center}
\end{figure}
\begin{figure}[htbp]
  \begin{center}
\includegraphics[width=0.93\linewidth]{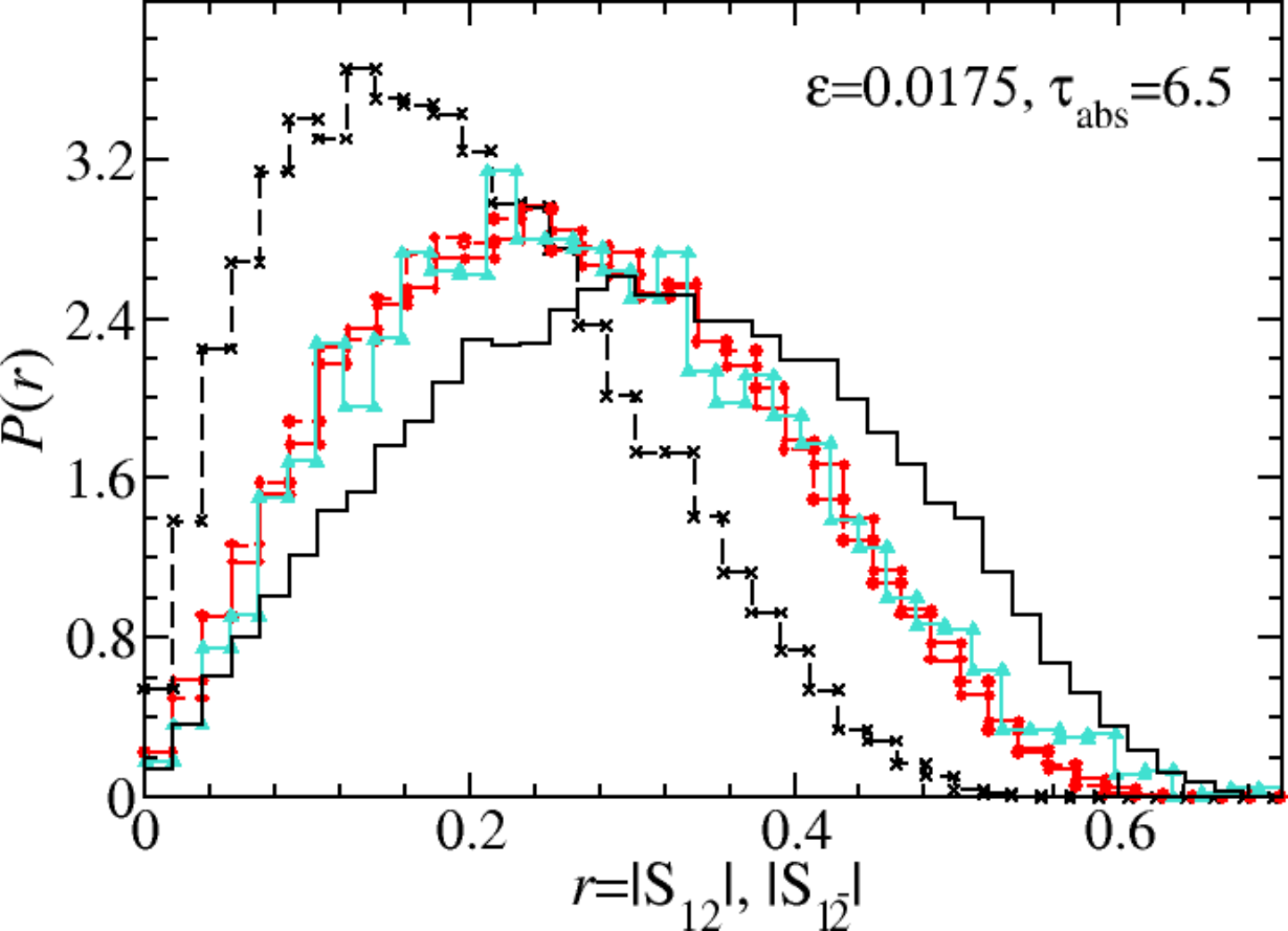}
\includegraphics[width=0.9\linewidth]{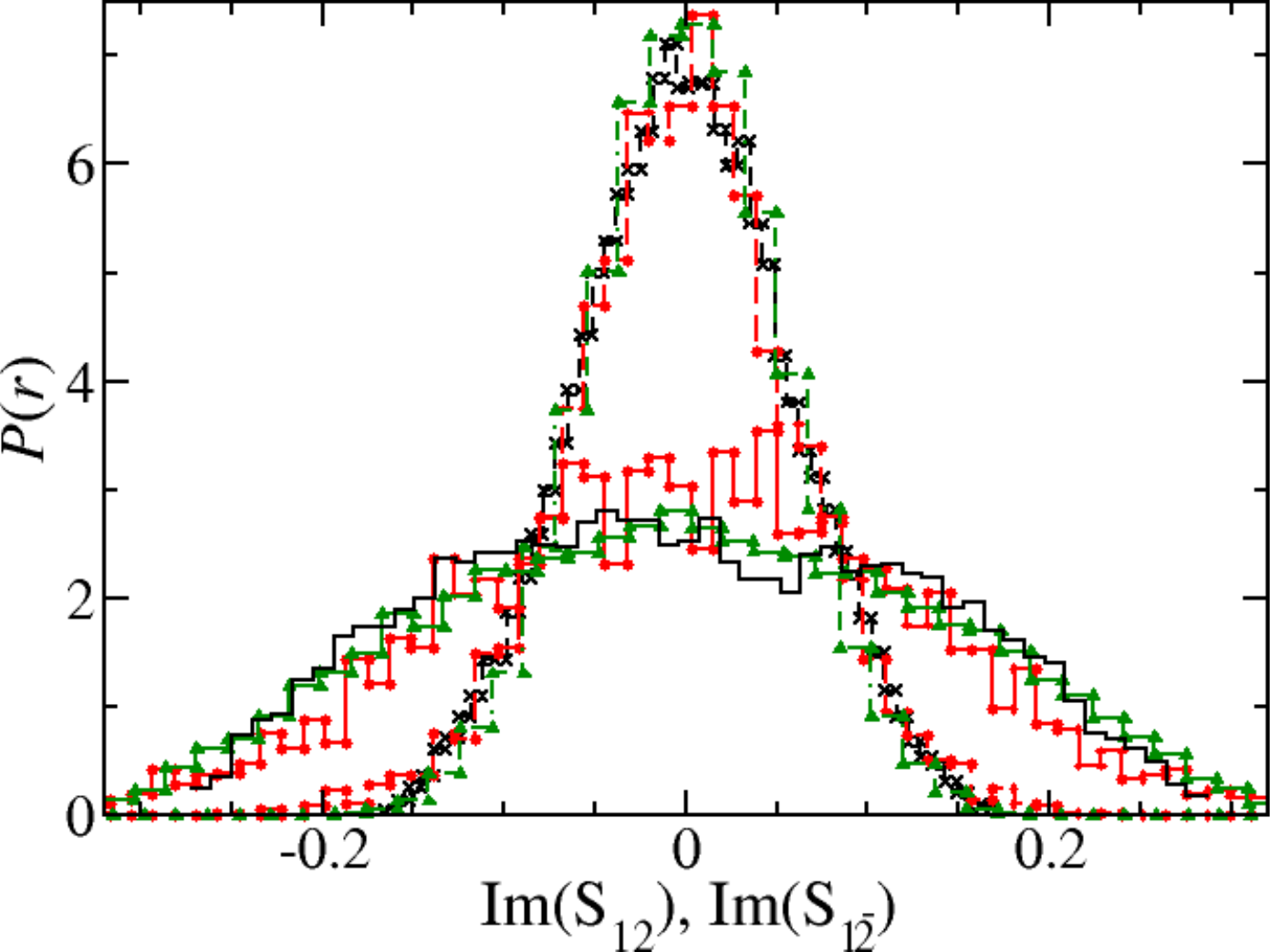}
\caption{Top: Modulus of $S_{12}$ and $S_{1\bar 2}$ (solid lines and dashed lines) for the cases with two coupling bonds (black solid line and dashed line with crosses) and four coupling bonds (red lines with dots). For the case with four coupling cables the distributions are well described by RMT (turquoise line with triangles). Bottom:  Comparison of the experimental distributions for the same experimental data as in~\reffig{Fig4} for the imaginary parts of $S_{12}$ and $S_{1\bar 2}$ (black solid line and dashed line with crosses) with numerical results for the corresponding quantum graph (green solid and dashed lines with triangles) and RMT simulations (red solid and dashed lines with dots) obtained by inserting into the RMT $S$ matrix a GSE matrix of the form~\refeq{Ham} where in $V$ all matrix elements were set to zero except the elements $(i_0,\bar j),(\bar i_0,j),(\bar j,i_0),(j,\bar i_0)$ for 5 values of $i_0$ and $j=1,\dots ,N$ with $N=200$ in the $2N\times 2N$-dimensional Hamiltonian $\hat H$,~\refeq{Ham}, for an ensemble of 300 $S$ matrices and $\tau_{\rm abs}=25$.
          }\label{Fig9}
  \end{center}
\end{figure}

\section{Distributions of the $S$-matrix elements\label{SecDistr}}
In the top panel of~\reffig{Fig4} the distributions of the $S$-matrix elements $S_{12}$ and $S_{1\bar 2}$ are compared. They are obtained from measured transmission spectra from $P_2$ to $P_1$, denoted by $S_{12}$, and from $P_{\bar 2}$ to $P_1$ denoted by $S_{1\bar 2}$, respectively. In the middle and bottom panel these distributions are compared to those obtained for the numerically computed $S$ matrix. Further results are shown in~\reffig{Fig5}, also for the distribution of $S_{11}$ for another size of absorption. The agreement is very good. Hence, we may conclude that the experiments indeed may be used to model the corresponding symplectic graph, implying that we may use numerical computations as ancillary data for a better understanding of experimental ones. Indeed, the experimental results were at first a surprise, because contrary to the property, that the distributions of all off-diagonal elements of the $S$ matrices of typical quantum-chaotic scattering systems with orthogonal or unitary symmetry -- like GOE and GUE graphs --, respectively, are the same and coincide with RMT predictions~\cite{Dietz2010,Hul2012,Lawniczak2020,Zhang2022,Kumar2013,Pluhar2014,Kumar2017}, those for $S_{12}$ and $S_{1\bar 2}$ clearly deviate from each other. The distributions are broader for scattering within a block, $S_{12}$, and sharper peaked around zero for $S_{1\bar 2}$, implying that scattering between the spin-up and spin-down system is less likely than spin-preserving scattering. This observation led us to the conclusion that the discrepancies originate from the fact that in the symplectic graph and microwave network only two of the nine vertices in one GUE graph are coupled to the corresponding two in the other GUE graph. To corroborate this, we also computed the $S$ matrix of the symplectic graph in which the subgraphs are coupled by four bonds, whose spectral properties are exhibited in~\reffig{Fig2}. Results for the distributions of the $S$-matrix elements are shown in Figs.~\ref{Fig6}-\ref{Fig7}. For the graph with four coupling bonds the distributions of $S_{12}$ and $S_{1\bar 2}$ agree well. Generally, for $a\ne b$ the distributions for the real and imaginary parts lie on top of each other for both cases, however the distributions for the case with two coupling bonds clearly deviate from those for four coupling bonds. On the contrary, for $a=b$ the distributions for the real and imaginary parts differ for each case, whereas the distributions of the two cases lie on top of each other. 

We also compared the experimental and numerical results for the $S$-matrix distributions with RMT predictions derived on the basis of the Heidelberg approach employing Hermitean random matrices from the GSE. As outlined in the following section,~\refsec{SecRMT}, agreement with RMT predictions is only found for the distributions of $S_{aa}$~\cite{Lawniczak2023}, whereas deviations are observed for $S_{ab} \, (a\neq b)$ for the case with two coupling bonds. Note that the matrix elements $S_{a\bar a}$ vanish~\cite{Rehemanjiang2016,Lu2020}. 

\section{Random-matrix simulations\label{SecRMT}}

The RMT simulations were performed using the $S$-matrix formalism~\cite{Mahaux1969}, that is,~\refeq{Sab} in the main text~\cite{Verbaarschot1985,Fyodorov2005,Dietz2009,Dietz2010,Kumar2013,Kumar2017}. In the present case there are $\tilde M=2M=4$ leads. The random matrix $H$ appearing in~\refeq{Sab} is drawn from the GSE, given in~\refeq{Ham}. The elements of the matrix $W$ describe the coupling of the modes in the microwave networks to the environment via the antenna ports. Furthermore, $\Lambda$ fictitious channels, that account for absorption~\cite{Dietz2009,Dietz2010} are introduced in each GUE graph. Hence altogether, there are $2(\Lambda+2)$ channels. To ensure that the $2(\Lambda +2)\times 2N$-dimensional coupling matrix $W$ complies with the symplectic properties of $H$, we diagonalized a random matrix from the GSE and used $\Lambda +2 < N$ of its pairs of eigenvectors associated with degenerate eigenvalues to generate $W$. Then the orthogonality property holds for $W$, $W_c^\dag W_d=\gamma_c \delta_{cd}\mathds{1}_2/\pi$. This is in accordance with the property that the frequency-averaged $S$ matrix is diagonal in all the microwave-network experiments. The quantities $\gamma_c$ are the input parameters of the RMT model. They are related to the transmission coefficients as
\begin{equation}
T_c =4\frac{\gamma_c}{(1+\gamma_c)^2}.
\label{Transm}
\end{equation}
For $c=1,\bar 1,2,\bar 2$ the transmission coefficients are determined from measurements of the reflection spectra $S_{aa}$ via the relation $T_a = 1 - \vert\overline{S_{aa}}\vert^2$~\cite{Dietz2010} with $\overline{X}$ denoting the spectral average of $X$. The transmission coefficients associated with the fictitious channels are all set to $T_f$. Here we chose $\Lambda=25$ in the random-matrix simulations. These are related to the absorption strength $\tau_{\rm abs}$ through the Weisskopf relation~\cite{Blatt1952,Dietz2009}, $\tau_{\rm abs}=2\Lambda T_{f}$. This parameter is used as fit parameter and is determined by comparing the RMT results for the distributions of the $S$-matrix elements, to the experimental ones. The resulting distributions for $S_{12}$ and $S_{1\bar 2}$ are equal. In~\reffig{Fig8} the RMT results for the distributions of the modulus of $S_{11}$ (top) and of $S_{12}$ (bottom) are compared to those obtained from experiments and numerical computations. In the experiments absorption was varied as indicated in the legends of the panels by adding attenuators to corresponding bonds of the subgraphs. The agreement is quite good for that case, whereas the RMT curves clearly deviate from those for the microwave network and quantum graph for $S_{1\bar 2}$, as can be inferred from the middle part of ~\reffig{Fig7}. Good agreement is only found for the case with four coupling cables as demonstrated for the modulus of $S_{12}$ and $S_{1\bar 2}$, respectively, in the top panel of~\reffig{Fig9}. 

As mentioned above, we suspect that the discrepancies between the distributions for $S_{12}$ and $S_{1\bar 2}$ observed in the experiments and the random-matrix simulations based on the scattering matrix formalism~\refeq{Sab} originate from the fact that in the RMT model Hermitean matrices from the GSE of the form~\refeq{Ham} with a full $V$ matrix are used. This corresponds to a coupling between all matrix elements of $H_0$ and $H_0^\ast$. However, in the microwave network and quantum graph, only two vertices of one subgraph are connected to the analogue ones in the other subgraph. Accordingly, we modified the Hamiltonian~\refeq{Ham} entering the $S$ matrix, and set all entries of the matrix $V$ to zero except the elements with indices $(i_0,\bar j),(\bar i_0,j),(\bar j,i_0),(j,\bar i_0)$ for 5 values of $i_0$ and $j=1,\dots ,N$ with $N=200$ in the $2N\times 2N$-dimensional Hamilton matrix $H$,~\refeq{Ham}, that is, drastically reduced its rank.  As demonstrated in the lower part of~\reffig{Fig9} then the agreement between the results from the RMT simulations for an ensemble of 300 $S$ matrices with $\tau_{\rm{abs}}=25$, the experimental ones, and the numerical ones  is good.

\section{Conclusions\label{SecConcl}}
We performed experiments with open microwave networks with symplectic symmetry, and analyzed distributions of the off-diagonal matrix elements $S_{ab}$ (which is equivalent to $S_{\bar a\bar b}$) and $S_{a\bar b}$ (which is equivalent to $S_{\bar a b}$) with $a\ne b$. We find that their distributions differ, in contradiction to RMT predictions. The discrepancies are attributed to the fact that the two subgraphs, that form the symplectic graphs, are coupled by just one pair of coaxial cables. This is confirmed with numerical computations for symplectic graphs, in which the subgraphs are coupled by 2 or more pairs of cables. Furthermore, we performed RMT simulations with a Hamiltonian of the form~\refeq{Ham}, where the matrix $V$ has a rank smaller than its dimension, corroborating our conjecture. Our findings imply that strictly speaking, because of the sparsity of the matrix $V$, the microwave networks employed in the present work do not mimick quantum-chaotic $S$ matrices described by the scattering formalism~\refeq{Sab} with $H$ a random matrix~\refeq{Ham} from the GSE, even though their spectral properties agree well with GSE statistics. Yet, the properties of the $S$ matrix are well described by a random-matrix model of the form~\refeq{Sab} with $V$ in~\refeq{Ham} a matrix of rank lower than its dimension. Thus, the $S$ matrices of such microwave networks and of the corresponding quantum graphs exhibit universal properties. Currently, we derive analytical expressions for the distributions of the off-diagonal $S$-matrix elements of random matrices belonging to the symplectic universality class, composed of fully connected subsystems $H_0$ and $H_0^\ast$, i.e., random matrices from the GSE, which already is quite cumbersome. The next step will be to modify the model such that the analytical derivation using a similar approach becomes feasible for the case of partially connected subsystems, which is challenging.

\begin{acknowledgements}
	This research was funded by the Deutsche Forschungsgemeinschaft (DFG, German Research Foundation) within the project Stochastic Quantum Scattering -- New Tools, New Aspects, DFG project number 540160740. Two of us (JC and BD) acknowledge financial support for the experiments from the China National Science Foundation (NSF), grants~11775100, 12247101 and 11961131009. One of us (BD) is grateful for funding from the Institute for Basic Science in Korea, project IBS--R024--D1. 
\end{acknowledgements}
\bibliography{Refs_GSE_DFG_Projekt,References_GSE_Paper}
\end{document}